\documentclass[11pt,twoside]{article}

\usepackage{asp2006}
\usepackage{epsf}
\usepackage{lscape}

\begin{document}
\title{CME on CME interaction on January  17,  2005}   
\author{C. Bouratzis, P. Preka Papadima, X. Moussas, A. E. Hillaris, C. Caroubalos}   
\affil{University of Athens, Zografos (Athens) , GR-15783, Greece}    
\author{C. E. Alissandrakis}
\affil{University of Ioannina, GR-45110 Ioannina, Greece}  
\author{P.  Tsitsipis, A. Kontogeorgos}
\affil{Technological Educational Institute of Lamia, Lamia , GR-35100, Greece}   

\begin{abstract} 
On January 17, 2005 a complex radio event associated with an X3.8 SXR flare and two fast Halo CMEs 
(CME$_1$ \& CME$_2$ henceforward) in close succession was observed.  We present combined 
ARTEMIS--IV \& WIND WAVES dynamic spectra which provide a complete view of the radio emission 
induced by shock waves and electron beams from the low corona to about 1 
A.U. These are supplemented with data, from the Nan\c cay Radioheliograph (NRH), GOES, 
EIT and LASCO for the study of the associated flare and CME activity.
\end{abstract}
\section*{Overview of the Event--Origin of the Radio Signatures}    
The complex event on  January  17,~2005~was observed with  the  
radio-spectrograph ARTEMIS-IV \citep{Caroubalos01}, the Nan\c cay Radioheliograph \citep[NRH][]{Kerdraon97},
the WIND WAVES \citep{Bougeret95} and the GOES, EIT and LASCO \citep{Yashiro04}. 
The most important radio bursts were two Type IIIGGs, two type II shocks and a long lasting type IV continuum
which covered the ARTEMIS-IV/WIND/WAVES spectral range for hours; the EIT \& NRH images
indicate that they all originated near AR 720.

\begin{figure}[!ht] 
\plotone{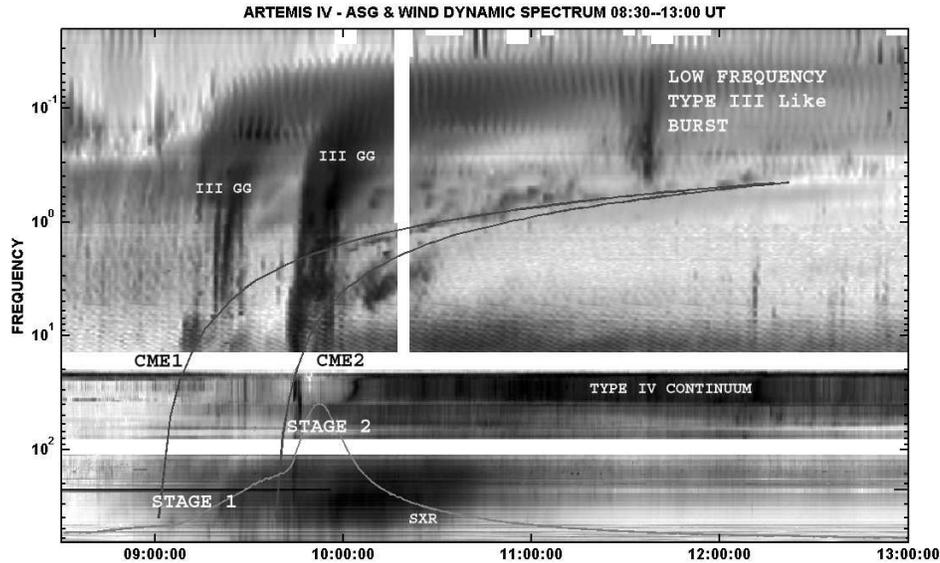}
\caption{WIND \& ARTEMIS IV Dynamic Spectrum, GOES SXR profile \& the frequency--time plots of CME$_1$ \& CME$_2$.
The type IV continuum the  two type III GG bursts, orginated at the two
stages of the SXR flux rise, and the type III like burst at the convergence of the 
CME$_1$ \& CME$_2$ fronts are duly annotated on the plot.}
\label{F3}
\end{figure}

The GOES records report an X3.8 SXR flare from 06:59 UT to 10:07 UT, with maximum at 09:52 UT. 
The halo CME$_1$ was first recorded by LASCO at 09:30:05 UT; it was launched around 09:00:47 UT. 
The next CME$_2$ was first recorded at 09:54 UT and was launched around 09:38:25 UT; it was found to 
overtake CME$_1$ at about 12:45 UT at a height of approximate 37 solar radii. The interaction of two 
fast CMEs (the speed of CME$_1$ was 2094 km$/$sec \& the speed of CME$_2$ 2547 km$/$sec)
is at variance with previous works \citep{Gopalswamy01, Lehtinen05} that focus on fast--slow CME interaction.

The type II shocks were recorded by the WIND/WAVES; their  extensions into the high frequencies 
are probably masked within the type IV continuum (Fig. \ref{F3}).  
The frequency–-time plots of CME$_1$ and CME$_2$ fronts were calculated from the 
LASCO Height vs. Time Plots and the Hybrid Coronal Density Model of \citet{Vrsnak04}; they are well
associated  with the type II lanes and are thus interpreted both as \textit{CME Bow Shocks}. 
As they approach each other (at about 11:37 UT) a low frequency type III-like burst is recorded, 
probably tracing shock accelerated energetic electrons. The CME$_1$ \& CME$_2$ lift--off is 
well associated, in time, with the two \textit{stages} of the  X3.8 SXR flux rise.

The two type III groups (08:41--09:42 UT), lastly, are also
well associated with the two \textit{stages} of the  X3.8 SXR flux rise, therefore appears that their 
exciter electrons originated in the low corona during the impulsive phase of the flare.


\end{document}